# Origin of a Superlattice Observed in $Li_{0.9}Mo_6O_{17}$ by Scanning Tunneling Microscopy

Ling Fu[1], Aaron M. Kraft[1], Martha Greenblatt[2], M. C. Boyer[1*]

[1]Department of Physics, Clark University, Worcester, MA 01610, USA

[2]Department of Chemistry and Chemical Biology, Rutgers University, Piscataway, NJ 08854, USA

**Abstract:**

We use scanning tunneling microscopy to study the lithium purple bronze ($Li_{0.9}Mo_6O_{17}$) at room temperature. Our measurements allow us to identify the single-crystal cleave plane and show that it is possible to obtain clean cleaved surfaces reflecting the crystal structure without the complications of nanoscale surface disorder. In addition to the crystal lattice, we observe a coexisting discommensurate superlattice with wavevectors $\boldsymbol{q} = 0.5\boldsymbol{a^*} \pm 0.25\boldsymbol{b^*}$. We propose that the origin of the superstructure is a surface reconstruction which is driven by cleaving along a crystal plane which contains in-plane $MoO_4$ tetrahedra connected to out-of-plane $MoO_6$ octahedra through corner-sharing oxygens. When combined with spectroscopic measurements, our studies show a promising avenue through which to study the complex physics within $Li_{0.9}Mo_6O_{17}$.

**Introduction:**

The lithium molybdenum purple bronze ($Li_{0.9}Mo_6O_{17}$) has garnered considerable interest due to a wealth of interesting physics contained within the low-dimensional material. Bulk and surface measurements on $Li_{0.9}Mo_6O_{17}$ suggest that the extreme one-dimensional nature may lead to non-Fermi liquid behavior and the realization of Luttinger Liquid physics from low temperatures up to room-temperature.[1-4] Resistivity measurements indicate that $Li_{0.9}Mo_6O_{17}$ undergoes a metal-to-insulator transition [1] near 25 K, though the origin of this transition is still a matter of debate. Suggested origins for this transition include charge or spin density waves [5-7], a dimensional crossover [6,8], and disorder-induced carrier localization [9-11]. In addition to the 25 K transition, $Li_{0.9}Mo_6O_{17}$ enters a superconducting state below ~2 K.[12] The unconventional superconducting state arising within the quasi-one dimensional material is three dimensional and proposed to be spin triplet.[13,14] How this three-dimensional superconducting state emerges within the lower dimensional material and the nature of electron pairing are open questions.

---

* To whom correspondence should be addressed: mboyer@clarku.edu

Scanning tunneling microscopy (STM) can be used to provide nanoscale details of temperature-induced phase transitions [15,16] and can give details regarding the symmetry of the superconducting order parameter, [17-19] which makes STM a possibly important tool in understanding the complex physics within $Li_{0.9}Mo_6O_{17}$. However, previous STM measurements on $Li_{0.9}Mo_6O_{17}$ have observed nanoscale surface disorder [4,20] which can lead to measured effects which differ significantly from the bulk. Our STM measurements on $Li_{0.9}Mo_6O_{17}$ indicate that surface disorder is not intrinsic to cleaved crystals, opening numerous avenues for using STM to explore the near-25 K transition and nature of the superconducting state in $Li_{0.9}Mo_6O_{17}$.

**Purple Bronzes:**

The lithium molybdenum purple bronze, $Li_{0.9}Mo_6O_{17}$, is part of a larger family of purple bronzes given by the chemical formula $A_{0.9}Mo_6O_{17}$ where A = Na, K, Tl, or Li. While there are slight symmetry differences among the Na, K, and Tl purple bronzes, their crystal structures generally consist of four layers of corner-sharing $MoO_6$ octahedra stacked along the *c*-axis sandwiched between single layers of $MoO_4$ tetrahedra which corner share with the neighboring $MoO_6$ octahdra.[21-24] These $MoO_4$ tetrahedral layers are separated from one another by a single layer of Na, K, or Tl ions (**Figure 1a**). Temperature-dependent resistivity measurements show that Na, K, and Tl purple bronzes are quasi-two dimensional metallic compounds at room temperature with in-plane resistivity several orders of magnitude smaller than the out-of-plane resistivity, as well as a low-temperature upturn near 100 K indicating the onset of a charge-density-wave state.[25-27]

The crystal structure of $Li_{0.9}Mo_6O_{17}$ (**Figure 1b**) is also comprised of corner-sharing $MoO_6$ octaherdra and $MoO_4$ tetrahedra but in a somewhat different configuration than the other three purple bronzes. Instead of a slab composed of four layers of corner sharing $MoO_6$ octahedra, there are three layers. Instead of terminal $MoO_4$ tetrahedral layers at either ends of the octahedral blocks separated by Li ions, corner-sharing polyhedra continue across the Li ion layers.[28] These structural differences result in six distinct Mo sites, typically identified by number, within the unit cell of $Li_{0.9}Mo_6O_{17}$ compared to only three in each of the Na, K, and Tl purple bronzes. The $Li_{0.9}Mo_6O_{17}$ crystal structure is monoclinic (space group $P2_1/m$) with room-temperature lattice parameters determined by neutron scattering to be $a = 9.4909$ Å, $b = 5.5239$ Å, $c = 12.7530$ Å, $\beta = 90.588°$.[29] In this manuscript we will use the axis convention used by McCarroll et al. in reference [30]; other publications utlize a rotated crystal coordinate system which essentially, up to a sign, interchanges the *a*- and *c*-axes.

Resistivity measurements indicate that $Li_{0.9}Mo_6O_{17}$, unlike the other three purple bronzes, is quasi-one dimensional where the in-plane resistivity along the a-axis is ~100 times larger than that along the b-axis.[1] Conduction electrons are restricted to flow along Mo1-Mo4 zigzag chains.[28,31] The crystallographic structural differences, and hence resulting dimensionality differences, are likely attributed to the small ionic size of the Li ion.[28] Similar to temperature-dependent measurements on the three other purple bronzes, $Li_{0.9}Mo_6O_{17}$ has an upturn in resistivity though, at the much lower temperature of ~25 K.[1] While this upturn in the other compounds is attributed to the onset of a charge-density-wave state, the origin of this upturn in $Li_{0.9}Mo_6O_{17}$ is a matter of debate.

STM has been successfully employed to image the Na and K purple bronzes at room temperature.[32,33] In each study, STM is able to resolve the $MoO_4$ units of the terminating $MoO_4$ tetrahedral layer. The lack of corner sharing polyhedra across the Na and K layers makes for a natural cleave plane as indicated by the dotted line in Figure 1a. The Na and K ions which lie along the cleave plane are not observed by STM, as in these materials the ions donate their valence electron to the conduction band. In addition to the crystal structure, STM topographies taken on $K_{0.9}Mo_6O_{17}$ below the charge-density-wave transition temperature show a 2x2 superlattice superimposed on the crystal lattice, a direct imaging of the charge-density-wave state.[33] On the other hand, previous STM topographies acquired on $Li_{0.9}Mo_6O_{17}$ show hints of 1-dimensional chains, but generally show disordered surfaces with no clear crystal structure.[4] Such disordered surfaces may not be unexpected given that in the $Li_{0.9}Mo_6O_{17}$ crystal structure $MoO_4$ tetrahedra connect $MoO_6$ octahdra across the Li ion layer resulting in no clear crystal cleave plane. Given that surface disorder can lead to surface effects which differ from the bulk, the ability to obtain clean crystal surfaces in $Li_{0.9}Mo_6O_{17}$ would allow for STM to probe properties relevant to the bulk material physics without the complications of surface disorder.

**Experiments and Results:**

Single crystal $Li_{0.9}Mo_6O_{17}$ samples were grown using a temperature gradient flux technique described in detail in reference [30]. STM studies were conducted using an RHK Technology PanScan STM at a room-temperature in ambient conditions using a mechanically sharpened Pt-Ir tip. Samples were cleaved by mechanically hitting a cleaver bar epoxied to the sample surface which is expected to expose the *a-b* crystal plane, [1,4] and then were inserted into the STM.

Topographic images acquired on $Li_{0.9}Mo_6O_{17}$ showed some surface structure variation and we did observe nanoscale disordered surfaces with 1-dimensional chain signatures (**Figure 2a**). However, the most common surface we observed was one with clear molecular resolution and where nanoscale disorder is absent (**Figure 2b**). The Fourier transform (FFT) of this topography evinces the crystallographic structural Bragg peaks (**Figure 2c**). We are able to identify the ***a*** and ***b*** reciprocal lattice vectors confirming that the *a-b* crystal plane was exposed on cleaving. In addition to the structural Bragg peaks, we identify additional topographic periodicities in the FFT indicated by yellow arrows, the origin of which is not immediately obvious.

To determine the specific cleave plane within the unit cell of $Li_{0.9}Mo_6O_{17}$, we use Fourier filtering to isolate the crystal lattice and remove any extra periodicities from the image (**Figure 2d**). In the filtered image, we see alternating bright and dark chains running along the *b*-axis and that the dark chains are slightly off center from the midpoint between adjacent bright chains. Comparison to the $Li_{0.9}Mo_6O_{17}$ crystal structure allows us to determine that cleaving exposes the Mo3 tetrahedra – Mo5 octahedra plane (**Figure 2e**). Further, we identify the bright chains as the Mo5 octahedra and the darker chains as the Mo3 tetrahedra. We make this determination based on calculations of the partial local density of states for Mo atoms by Nuss et. al.[34] These calculations find that the Mo5 atoms have a higher density of states near the Fermi energy than do the Mo3s; this leads to the Mo5s appearing brighter in our STM images.

Next, we turn our attention to the additional observed topographic periodicities, which we term the superlattice. Careful analysis of the topography shows that the superlattice is comprised of "bright spots" which are tied to, but are slightly offset from, the Mo5 locations of the crystal lattice. The wavevectors associated with the periodicities can be extracted from the Fourier transform of the topography and are determined to be ***q*** = 0.5***a**** ± 0.25***b****. The diffuse nature of the superlattice peaks in the Fourier transform is due to the discommensurate nature of the superlattice which we observe in our images and which can be seen visibly in Figure 2b.

To further understand the characteristics of the superlattice, we extract linecuts through the Mo5 sites of the topography along the *a*- and *b*- crystal axes (**Figure 3**). The linecut along the *b*-crystal axis shows a large peak at the superlattice site at the position expected for a Mo5 octahedron; there is no evidence for duplicate peaks near the superlattice site nor is there a measured spatial shift in the peak location in comparison to that of the expected Mo5 octahedron location.

Along the *a*-axis, locally the bright spots of the superlattice appear in alternating Mo5 rows. For comparison, we examine the topographic profile along a Mo5 row with the superlattice maxima to a neighboring row without. Within a locally commensurate superlattice region, the linecut shows a single peak at the superlattice affected site. This peak is both wider and higher than the standard Mo5 crystal lattice peak. In addition, the peak location is shifted $0.43 \pm 0.07$ Å along the *a*-axis in the direction of the nearest neighboring Mo3 tetrahedral chain.

We acquired spectroscopic measurements on the $Li_{0.9}Mo_6O_{17}$ surface. **Figure 4** shows the numerical derivative of an average of ~1000 current vs. voltage measurements. This spectrum, which represents the convolution of the sample density of states with the derivative of the Fermi function evaluated at room temperature, reflects bulk electronic properties of $Li_{0.9}Mo_6O_{17}$. The high-energy features seen in the spectrum are consistent with those found in ab initio-calculated density of states (reference [34]) when we thermally broaden the calculated density of states to 295 K.

**Discussion:**

In addition to the crystal lattice, we observe a superlattice in the Mo3-Mo5 crystal plane which is tied to the Mo5 octahedral locations. Interestingly, the superlattice is two dimensional within this quasi-one dimensional compound. Furthermore, this superlattice is observed at room temperature, well above the 25 K transition, making it unlikely that the origin of the superlattice is tied to any theories explaining the low-temperature transition.

To help determine the nature of the origin of the superlattice, we acquired topographies over a range of sample biases. We find the observed superlattice is unchanged over the wide range of positive (+10 mV to +250 mV) and negative (-10 mV to -800 mV) sample biases for which we acquired topographies. That there are no detected differences in the superlattice when probing filled and empty sample states over a wide bias range suggests that the superlattice may be structural in origin. In comparing the characteristics of the superlattice to the crystal structure of $Li_{0.9}Mo_6O_{17}$, we show that the superlattice features can be well understood in terms of a surface reconstruction. This surface reconstruction is driven by cleaving the crystal through a plane across which Mo-O polyhedra connect to one another via corner-sharing oxygens.

We have determined that cleaving $Li_{0.9}Mo_6O_{17}$ exposes the Mo3-Mo5 crystal plane. This indicates that the crystal cleave plane is between the Li ions (black dotted line in Figure 1b). This cleavage plane is directly through the Mo6 tetrahedra which are directly attached to Mo5 octahedra through corner-sharing oxygens. We clearly observe the Mo3 tetrahedra and Mo5 octahedra in our topographic images. However, we see no evidence for the Mo6 tetrahedra. For this reason, we determine that in regions such as seen in Figure 2b that the Mo6 tetrahedra are likely removed in the cleaving process. Because the Mo5 octahedra and Mo6 tetrahedra are coupled through corner-sharing oxygens, removing Mo6 tetrahedra will lead to a distortion of the Mo5 octahedra. In addition, removing a neighboring Mo6 tetrahedron will likely change the valence of the Mo5 octahedron. Both of these, in turn, are likely to lead to a surface reconstruction.

Our linecuts through Mo5 and Mo3 sites in the topography allow us to characterize this surface reconstruction. We find that the surface reconstruction involves a periodic shift of Mo5s along the *a*-axis by $0.43 \pm 0.07$ Å and no shift along the *b*-axis. In taking similar linecuts through the Mo3 sites along the *a*- and *b*-crystal axes, we find no evidence for a systematic shift of the Mo3 tetrahedra associated with the superlattice, though we note that the proximity of the shifted Mo5 octahedra toward neighboring Mo3 tetrahedra makes this analysis more difficult and may act to obscure small shifts. That we observe a shift for the Mo5 octahedra but not for the Mo3 tetrahedra makes sense in terms of the crystal structure and the effects of removing Mo6 tetrahedra; Mo3 tetrahedra will shift less because they are not directly connected to the Mo6 tetrahedra.

A surface reconstruction can explain not only the shifted peak in the linecut through the Mo5s along the *a*-axis, but also the increased height and width. In the $Li_{0.9}Mo_6O_{17}$ crystal structure, Li ions lie in the Mo3-Mo5 plane between neighboring Mo5 octahedra directly along the *a*-axis (Figures 2e and f). While we do not observe Li ions in the topography, since they donate their valence electron to the Mo1-Mo4 chains, their presence is noted in shift of Mo5 upward; the Mo5 octahedra are unable to purely shift along the *a*-axis but rather are forced higher by 23 ± 6 pm as they shift due to the neighboring Li ion. The width of the peak associated with a raised Mo5 octahedron as a result will appear wider as more of the peak is observed.

Finally, our topographic images show that the superlattice is discommensurate in nature; the superlattice is tied to the Mo5 sites with $\boldsymbol{q} = 0.5\boldsymbol{a}^* \pm 0.25\boldsymbol{b}^*$ but with noted phase shifts in the pattern. We note that this discommensurate nature differs between sample cleaves; for some sample cleaves we have detected a commensurate pattern without phase shifts over 100 Å while for others, such as in Figure 2b, we find much smaller commensurate domains. Such a dependence on sample cleaving further emphasizes that the superlattice is a surface phenomenon which does not exist within the bulk as well as prevents us from determining a characteristic commensurate domain size.

We have determined that the superlattice observed in our topographies likely originates from a surface reconstruction and have suggested why the superlattice is tied to the Mo5 sites. Our findings, that cleaving $Li_{0.9}Mo_6O_{17}$ leads to a surface reconstruction which involves a significant shift in Mo5 octahedra and minimal shift in Mo3 tetrehdra, are observed in preliminary density functional theory (DFT) calculations.[35] However, more detailed DFT calculations are needed to explain the origin of the specific pattern of the superlattice and the determined superlattice wavevectors. Given the large unit cell of $Li_{0.9}Mo_6O_{17}$ such calculations are computationally intensive and challenging, particularly without a detailed understanding of the fate of the corner-sharing Mo5-Mo6 oxygen atoms after cleaving.[35]

**Conclusions:**

Our topographic STM measurements on $Li_{0.9}Mo_6O_{17}$ show that it is possible to obtain clean crystal surfaces with sample cleaving. We have determined that cleaving the crystal exposes the Mo3-Mo5 *a-b* crystallographic plane. We observe a coexisting superstructure which we propose is structural in nature and whose origin is likely due to a surface reconstruction. If the Mo5 octahedral shifts associated with the superlattice we observe were to exist throughout the bulk, these shifts should be easily detected in x-ray and neutron scattering measurements. Our acquired spectroscopic measurements are consistent with the bulk electronic structure of Li$_{0.9}$Mo$_6$O$_{17}$. Overall, our work shows that STM can be used to study the atomic-scale details of Li$_{0.9}$Mo$_6$O$_{17}$ physics which will be particularly useful in low-temperature studies where STM can be used to study the 25 K transition as well as the superconducting state below 2 K without the complications of nanoscale disorder.

**Acknowledgements:**
We thank Markus Aichhorn, Gernot Kraberger, Jennifer E. Hoffman, and Anjan Soumyanarayanan for useful discussions and communications.

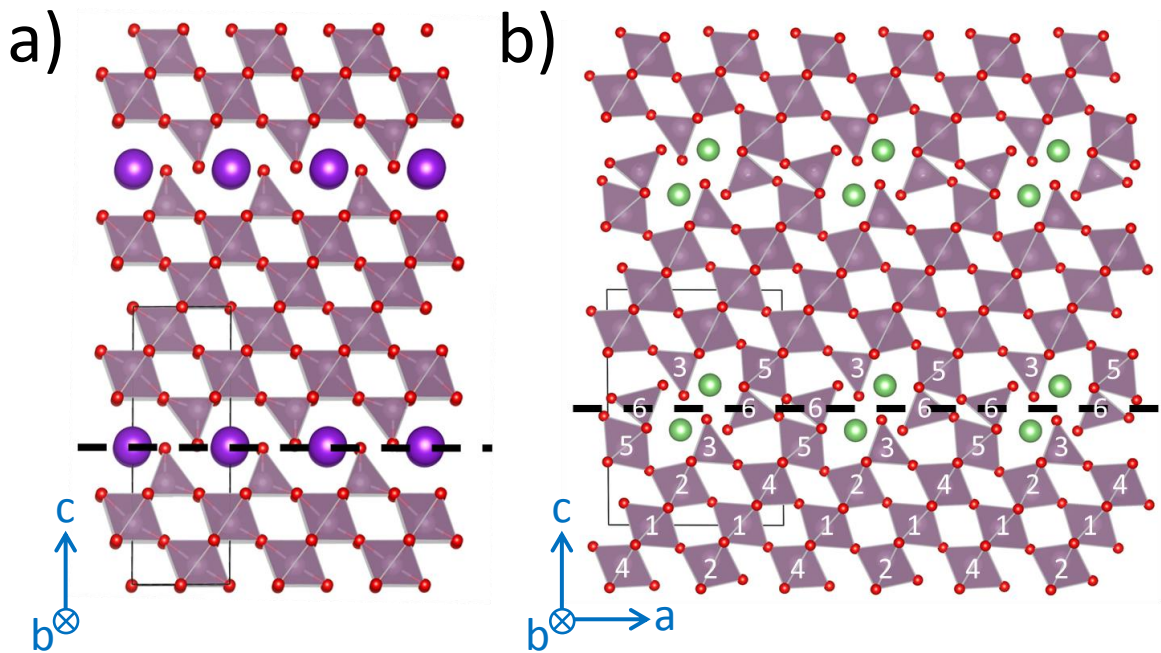

**Figure 1:** a) Crystal structure of $K_{0.9}Mo_6O_{17}$ based on neutron scattering data from [21]. The structure shows $MoO_6$ octahedra and $MoO_4$ tetrahedra. The polyhedra are connected through corner-sharing oxygens (seen in red). The dotted black line indicates the determined cleave plane. b) Crystal structure of $Li_{0.9}Mo_6O_{17}$ based on neutron scattering data from [29] showing Mo-O polyhedra. The six Mo sites are identified by number. The dotted black line is drawn through $MoO_4$ tetrahedra (identified as Mo6 tetrahedra) which connect Mo-O polyhedra through the Li ion layer. Both crystal structure images were drawn using Vesta [36].

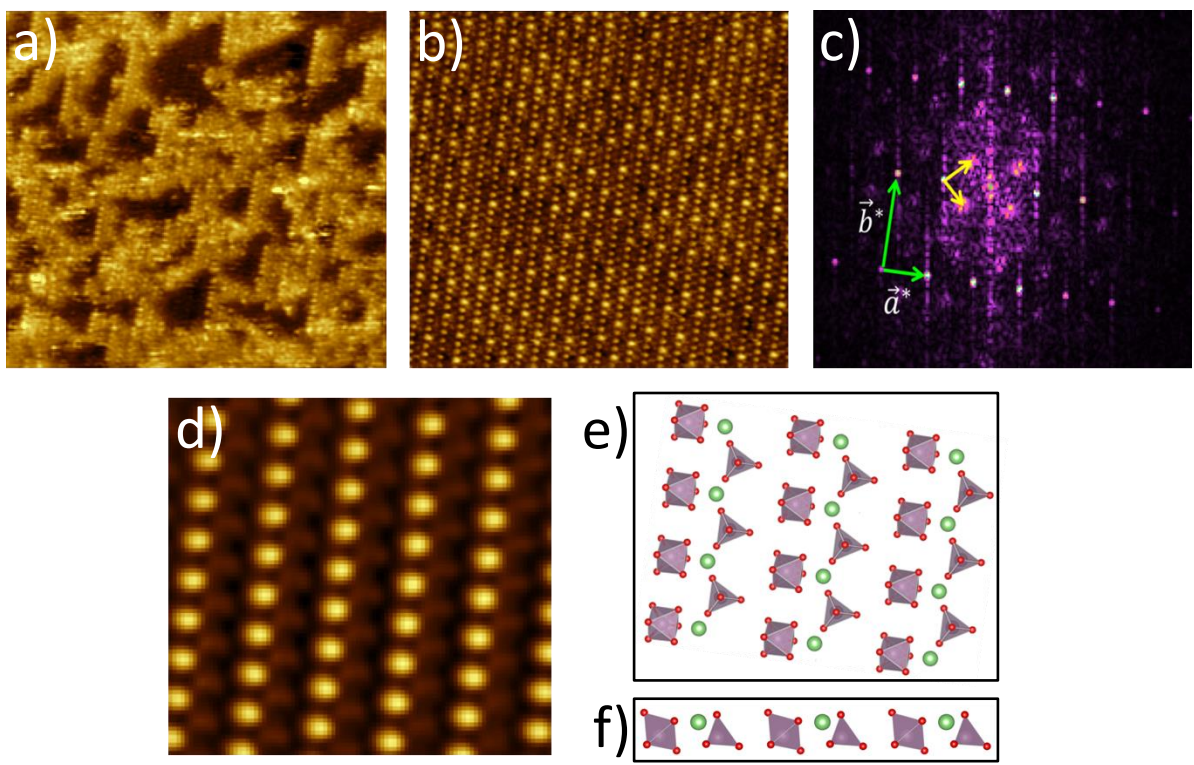

**Figure 2:** a) A 30 nm square scan ($V_{Sample}$ = +100 mV and I = 50 pA) of $Li_{0.9}Mo_6O_{17}$. While the topographic image shows nanoscale disorder, hints of one-dimensional chains are seen. b) 23 nm square topography ($V_{Sample}$ = -20 mV and I = -50 pA ) of the *a-b* crystal plane. Alternating bright and dark chains are seen running along the b-crystal axis. c) FFT of the image in (b). Reciprocal lattice vectors ***a*** and ***b*** are indicated by arrows in green. Wavevectors associated with a coexisting topographic superlattice are indicated by yellow arrows. d) Zoomed 5 nm image. Using Fourier filtering of the image in (b), the superlattice is removed and only the crystal lattice remains. Based on comparison to the crystal structure, seen in (e), and partial density of state calculations from reference [34] we identify the bright chains as Mo5 octahedra and dark chains as Mo3 tetrahedra. e) The Mo5 octahedral – Mo3 tetrahedral crystal plane of $Li_{0.9}Mo_6O_{17}$. Li ions appear in the same crystal plane as is emphasized by a side view of the crystal plane (f).

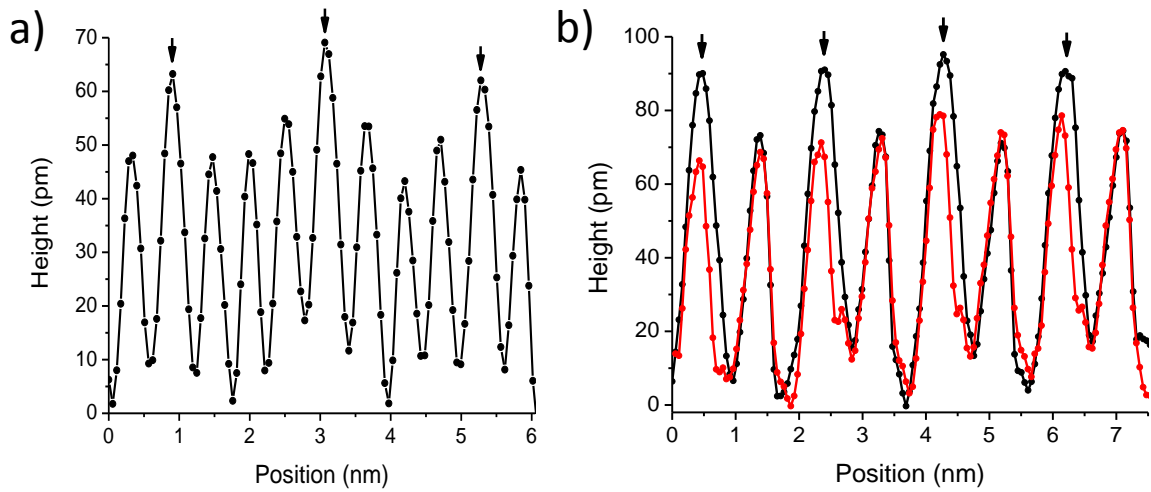

**Figure 3:** a) A linecut along the *b*-axis of $Li_{0.9}Mo_6O_{17}$ through Mo5 octahedra. The Mo5 superlattice sites are indicated with black arrows. There are no measured spatial shifts along the *b*-axis associated with the superlattice sites. b) Linecut along the *a*-axis of $Li_{0.9}Mo_6O_{17}$ through Mo5 octahedra. Black curve: Linecut through a Mo5 row with superlattice maxima. Red curve: Linecut through neighboring Mo5 row without superlattice maxima. At the superlattice sites (indicated by black arrows), the peaks are higher and shifted from the expected Mo5 octahedra locations.

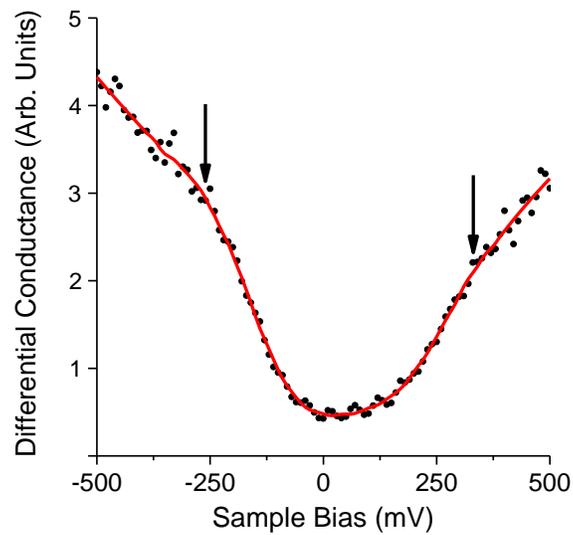

**Figure 4:** Plot of the differential conductance with sample bias. The high energy features indicated by arrows in the plot are consistent with the high energy features associated with bulk electronic properties of $Li_{0.9}Mo_6O_{17}$ as calculated in reference [34]. These spectral features are observed ubiquitously across different samples, sample cleaves, and tips.